\documentclass[twocolumn]{aastex63}

\usepackage{amsmath}
\usepackage{graphicx}

\newcommand{\msol}{M_\odot}

\newcommand{\E}[1]{\hspace{-0.01in}\times\hspace{-0.02in}10^{#1}}  
 
\newcommand{\angstrom}{\mbox{\normalfont\AA}}

\newcommand{\ha}{H\alpha}

\accepted{to the Astrophysical Journal}

\shortauthors{Medina et al.}

\graphicspath{{./}{figures/}}

\begin{document}

\title{Variability Timescales of $\ha$ on Active Mid-to-Late M dwarfs}

\correspondingauthor{Amber A. Medina}
\email{amber.medina@austin.utexas.edu}

\author[0000-0002-0786-7307]{Amber A. Medina}
\affiliation{Center for Astrophysics | Harvard \& Smithsonian, 60 Garden Street, Cambridge, MA~02138, USA}
\affiliation{Department of Astronomy, The University of Texas at Austin, Austin, TX 78712, USA}

\author[0000-0002-9003-484X]{David Charbonneau}
\affiliation{Center for Astrophysics | Harvard \& Smithsonian, 60 Garden Street, Cambridge, MA~02138, USA}

\author[0000-0001-6031-9513]{Jennifer G.\ Winters}
\affiliation{Center for Astrophysics | Harvard \& Smithsonian, 60 Garden Street, Cambridge, MA~02138, USA}

\author{Jonathan Irwin}
\affiliation{Center for Astrophysics | Harvard \& Smithsonian, 60 Garden Street, Cambridge, MA~02138, USA}

\author{Jessica Mink}
\affiliation{Center for Astrophysics | Harvard \& Smithsonian, 60 Garden Street, Cambridge, MA~02138, USA}



\begin{abstract}
We present a study of the variation timescales of the chromospheric activity indicator $\ha$ on a sample of 13 fully-convective, active mid-to-late M stars with masses between 0.1--0.3 solar masses. Our goal was to determine the dominant variability timescale and, by inference, a possible mechanism responsible for the variation. We gathered 10 or more high-resolution spectra each of 10 stars using the TRES spectrograph at times chosen to span all phases of stellar rotation, as determined from photometric data from the MEarth Observatories. All stars varied in their $\ha$ emission. For 9 of these stars, we found no correlation between $\ha$ and rotational phase, indicating that constant emission from fixed magnetic structures, such as starspots and plage, are unlikely to be the dominant source of $\ha$ emission variability. In contrast, one star, G 7-34, shows a clear relationship between $\ha$ and stellar rotational phase. Intriguingly, we found that this star is a member of the AB Doradus moving group and hence has the young age of 149 Myr. High-cadence spectroscopic observations of three additional stars revealed that they are variable on timescales ranging from 20--45 minutes, which we posit may be due to flaring behavior. For one star, GJ 1111, simultaneous \textit{TESS} photometry and spectroscopic monitoring show an increase in $\ha$ emission with increased photometric brightness. We conclude that low-energy flares are able to produce variation in $\ha$ on the timescales we observe and thus may be the dominant source of $\ha$ variability on active fully-convective M dwarfs.

\end{abstract}


\section{Introduction} \label{sec:intro}

Main-sequence stars with masses below 0.35$\msol$ are believed to be fully convective and lack the tachocline region that is thought to be essential to magnetic field generation models in partially convective stars. Fully convective stars nonetheless display a variety of magnetic phenomena like stellar spots, flares, and chromospheric emission. Chromospheric spectroscopic activity indicators like Calcium II H+K at 3933 $\angstrom$ and 3968 $\angstrom$ respectively and $\ha$ at 6562.8 $\angstrom$ are produced as plasma traveling along magnetic loops heats the chromosphere.  It is now well established that the emission is variable in time \citep[see][]{Suarez2016,Lee2010, Kruse2010}. By determining the timescales upon which the activity varies, we hope to gain insight to the physical mechanism responsible for generating magnetic fields in fully convective stars.

For partially convective stars, \citet{Suarez2018} and \citet{Fuhrmeister2019} showed that activity monitoring on day--week timescales can be used to determine rotation periods. The ability to measure rotation periods using this method implies the activity occurs predominantly from regions of roughly constant emission fixed to the stellar surface, perhaps originating from stellar spots or plage. Similarly, \citet{Shofer2019} used a sample of both inactive and active (M0-M9) dwarfs to examine periodicity in activity indicators for a sample of 154 stars with measured photometric rotation periods. Using a generalized Lomb-Scargle periodogram, they found that the rotation period is observed in at least one activity indicator, on only a minority of these stars, 66 of the 154. All of the rotation periods they measured are greater than one day as to avoid potential sampling aliases with their observing cadence. Only one of these stars, GJ 406, has a mass that puts it below the limit at which we expect the star to be fully convective. They noted that for the remaining stars the activity variation may be of a different origin.

\citet{Kruse2010} explored variability in $\ha$ on timescales less than an hour using observations of M0-M9 dwarfs from the Sloan Digital Sky Survey. They showed that variability in $\ha$ increased towards later spectral types, but with the majority of stars having three observations or less separated by 15 minutes, it is difficult to determine the timescale of the variations. In a related study, \cite{Lee2010} used high-cadence observations of mid-to-late M dwarfs observed over the course of one hour and found that fluctuations in $\ha$ likely occur on timescales of 30 minutes to an hour, and are less likely to occur on shorter timescales. They attributed these fluctuations to stellar flares which are known to enhance $\ha$ emission \citep{Kowalski2009,Kowalski2013,Fuhrmeister2018,DiMaio2020}. However \cite{Lee2010} and \citet{Kruse2010} did not explore how variability relates to stellar rotation.

In this study, we used spectroscopic and photometric observations of a sample of fully convective M dwarfs with masses ranging from 0.1--0.26$\msol$ to examine the timescale of the variability of $\ha$, and whether it is correlated with stellar rotation. Such a correlation may imply that the emission of $\ha$ is dominated by roughly constant emission from phenomena such as spots and plage that are fixed on the stellar surface and rotating into and out of view. A lack of correlation would indicate that another mechanism governs the variability. Understanding this timescale would allow us to better describe the stellar radiation environment, which is vital as atmospheric studies of planets around these stars will take place in the near future. Furthermore, understanding the various timescales associated with activity is highly relevant to radial velocity surveys because stellar activity can induce apparent radial velocity variations that mimic or mask the Keplerian signal of an orbiting exoplanet. In \S~\ref{sec:sample} we describe the stellar sample, followed by a description of the spectroscopic and photometric observations and data reduction in \S~\ref{sec:spec_obs}. We first investigated whether the variation in $\ha$ tracked stellar rotation, and we present those findings in \S~\ref{sec:methods}. We subsequently investigated the variation on much shorter timescales, and we present those findings in section \S~\ref{sec:HC}. We discuss our conclusions in \S~\ref{sec:conclusion}

\section{Stellar Sample}\label{sec:sample}

Our initial sample consisted of 10 single M dwarfs ranging from 0.1 to 0.26 solar masses. We selected these stars from a volume complete sample of 413 such M dwarfs that reside within 15 parsecs \citep{Winters2021}. \citet{Winters2021} determined the distances to these 10 stars using parallaxes from the second data release of \textit{Gaia} \citep{GaiaDR22018} and estimated masses using the $M_{K_{S}}$ - mass-luminosity relation presented in \citet{Benedict2016}. The mass estimates have a typical uncertainty of 4.7--14.0\% which are dominated by the scatter in the mass-luminosity relation.  All stars have been vetted for close companions as described in \citet{Winters2021}. These stars have rotation periods from 0.21 - 91.92 days \citep{Newton2016, Newton2018} determined from the periodic photometric modulation induced by star spots as measured by the MEarth survey \citep[][See Section
\ref{sec:mearth_obs}]{Nutzman2008, Irwin2015}. We chose our sample to span a range of rotation periods in each of two mass bins, namely a more massive bin with masses 0.24--0.26 $\msol$, and a less massive bin with masses 0.12--0.17$\msol$. Based on our findings for our initial sample of ten stars, we subsequently added three additional stars to our sample, GJ 1111, GJ 1167, and LEP 1431+7526 with masses ranging from 0.10--0.17 $\msol$. We present the stellar parameters for all 13 stars in Table \ref{tab:stellar_params}.

\begin{longrotatetable}
\begin{center}
\begin{deluxetable*}{lccccccccc} \label{tab:stellar_params}
\tablecaption{Stellar Parameters}
\tabletypesize{\footnotesize}
\tablehead{ 
\colhead{Star} & 
\colhead{RA} &
\colhead{DEC} & 
\colhead{Mass} &   
\colhead{Distance}&
\colhead{Rotation \tablenotemark{a}} &
\colhead{Semi-amplitude \tablenotemark{b}} &
\colhead{Semi-amplitude \tablenotemark{b}} &
\colhead{F-Test} &
\colhead{Confidence in}\\
\nocolhead{Star} & 
\nocolhead{RA} &
\nocolhead{DEC} & 
\nocolhead{Mass} &   
\nocolhead{Distance}&
\colhead{Period} &
\nocolhead{} &
\colhead{error} &
\colhead{Statistic} &
\colhead{F-Test Statistic \tablenotemark{c}}\\
\nocolhead{yup} & 
\colhead{hh:mm:ss.s} &
\colhead{dd:mm:ss} &
\colhead{M$_{\odot}$ }&
\colhead{pc} &
\colhead{days}&
\colhead{mag} &
\colhead{mag}&
\nocolhead{} &
\colhead{\%}} 
\startdata
&   &   &   &   &Low-mass Sample &  &  &\\\hline
GJ~170 & 04:30:25.2 & +39:51:00 & 0.17 & 10.89 & 0.72 & 0.0073 & 0.0015 & 1.12 & 57\\
G~97-15 & 05:04:14.8 & +11:03:23 & 0.15 & 10.19 & 0.84 & 0.0015 & 0.00019 & 0.48 & 34\\
LHS~1690 & 04:39:32.0 & +16:15:43 & 0.12 & 11.92 & 3.61 & 0.0037 & 0.0019 & 1.42  & 63\\
LHS~2686 & 13:10:12.7 & +47:45:19 & 0.15 & 12.19 & 28.54 & 0.0081 & 0.0011 &  1.70 & 68\\
LHS~1723 & 05:01:57.0 & -06:56:46 & 0.17 & 5.38 & 91.92 & 0.0056 & 0.00073 &  0.63 & 41\\\hline
&   &   &   &  &High-mass Sample &  &  &\\\hline
G~7-34 & 04:17:18.5 & +08:49:22 & 0.25 & 14.59 & 0.37 & 0.0111 & 0.00019 &  19.98 & 98\\
G~192-12B & 05:59:55.7 & +58:34:16 & 0.26 & 14.78 & 0.95 & 0.0041 & 0.00018 &  2.34 & 76\\
G~99-49 & 06:00:03.5 & +02:42:23 & 0.24 & 5.21 & 1.81 & 0.0020 & 0.00022 &  0.14 & 13\\
LTT~15516 & 18:42:45.0 & +13:54:17 & 0.25 & 10.94 & 8.06 & 0.0070 & 0.00075 &  1.81 & 69\\
LP~816-60 & 20:52:33.0 & -16:58:29 & 0.24 & 5.16 & 82.92 & 0.0061 & 0.00069 & 0.33 & 26\\\hline
&   &   &   &   &High-cadence Sample &  &  & &\\\hline
GJ~1167 & 13:09:31.3 & +28:58:42 & 0.17 & 5.46 & 0.22 & 0.008 & 0.00019 & -- & --\\
GJ~1111 & 08:29:49.5 & +26:46:34 & 0.10 & 3.58 & 0.46 & 0.001 & 0.00011 & -- & --\\
LEP~1431+7526 & 14:31:13.4 & +75:26:42 & 0.16 & 14.52 & 0.63  & 0.002 & 0.00069 & -- & --\\
\enddata
\tablenotetext{a}{Photometrically determined rotation period.}
\tablenotetext{b}{Semi-amplitude of rotational photometric modulation.}
\tablenotetext{c}{F-Test statistic which determines whether the model with $\ha$ varying in phase with stellar rotation is favored over a model where the $\ha$ variation is consistent with a straight line; see Equation 4.}
\end{deluxetable*}
\end{center}
\end{longrotatetable}

\section{Observations and Reductions}\label{sec:spec_obs}

\subsection{Spectroscopic Observations}
We obtained 10--17 high-resolution spectra of each of the ten targets in the original sample. We used the Tillinghast Reflector Echelle Spectrograph (TRES) located on the 1.5 meter telescope at the Fred Lawrence Whipple Observatory at Mount Hopkins in Arizona. TRES has a resolution of R $\approx$ 44,000 and covers the wavelength range 3900-9100 $\angstrom$. Exposure times ranged from 150s to 3~$\times$~1200s to reach a signal to noise ratio of 3-27 at 7150 $\angstrom$. The spectra were reduced using the standard TRES pipeline \citep{Buchhave2010}. For each star, we scheduled the observations to obtain full phase coverage of the rotational modulation. We acquired the observations from UT 2018 September 01 - December 31. After an initial analysis of the original sample,  we subsequently decided to observe GJ~1111, GJ~1167, and LEP 1431+7526. We gathered observations continuously of a single star for time spans of 3.94-8.37 hours as follows: we obtained four nights of high-cadence observations for GJ~1167 on UT 2019 March 06, April 03, May 04, May 05 for 4.37, 7.57, 6.03, and 5.94 hours per night respectively with exposures times of 500s. We obtained high-cadence observations of GJ~1111 on UT 2020 January 24 for 8.37 hours with exposure times of 600s. We obtained high-cadence observations of LEP~1431+7526 on UT 2020 February 02 for 3.94 hours with exposures times of 300s.

We measured the equivalent width (EWs) of $\ha$ according to the following equation,

\begin{equation}\label{eq:EW}
    \rm EW = \sum_{i=1}^{\rm N_{pix}} {\left( 1 - \frac{F_i(\lambda)}{F_c} \right)} \delta\lambda
\end{equation}

\noindent where F$_i(\lambda)$ is the flux per pixel included in the $\ha$ feature and F$_c$ is the average flux deduced from the adjacent continuum region. We summed the flux in each pixel including fractional pixels to measure the flux contained within  6560.3 $-$ 6565.3 $\angstrom$, which we define as the feature window. We measured the average flux value in continuum regions on either side of the $\ha$ feature. We defined the continuum regions from  6554.1 $-$ 6559.1 $\angstrom$ and 6566.5 $-$ 6570.5 $\angstrom$. In order to measure F$_c$, we first determined the average flux contained in each continuum region. F$_c$ is then the average of these two values.  We chose continuum regions as to avoid tellurics and complex molecular bands from the stellar photosphere. The uncertainty in the equivalent width is the product of the  measured EW value and the fractional uncertainties in F$_c$ and F$(\lambda)$ added in quadrature. Negative EW values denote emission. We provide the times of observation, EW values, and their uncertainties in Table \ref{tab:EWs}. 

\begin{deluxetable}{lccl}
\tabletypesize{\scriptsize}
\tablecaption{Catalog of Equivalent Widths \label{tab:EWs} (Table Format)}
\tablehead{ 
\colhead{Column} & 
\colhead{Format} &  
\colhead{Units} & 
\colhead{Description}}
\startdata 
1 & A10 & ... & Star Name \\
2 & F5.4 & BJD & Barycentric Julian Date \\
3 & F2.3 & $\angstrom$ & $\ha$ Equivalent Width\\
4 & F2.3 & $\angstrom$ & Error on $\ha$ Equivalent Width \\
\enddata 
\end{deluxetable}

\subsection{Photometric Observations}\label{sec:mearth_obs}
In order to determine if the variation in $\ha$ was correlated with the photometric rotational modulation, we photometrically monitored the first 10 stars in Table 1. We used MEarth North for eight of our targets and MEarth South for the remaining two targets with declinations below 0 degrees. The MEarth-North and MEarth-South arrays each consist of eight 0.4 meter robotic telescopes located on Mt. Hopkins in Arizona and at Cerro Tololo Inter-American Observatory (CTIO), Chile, respectively \citep{Nutzman2008,Irwin2015}. We obtained data on clear nights from UT 2018 September 01 - UT 2018 December 31 to overlap with the spectroscopic observations. In addition, we used all prior observations from the MEarth Survey of these targets. 

The photometric modulation caused by stellar spots rotating into and out of view allow us to measure rotation periods. We measured the rotation period for each target to ensure that we recovered the same value as that presented in \citet{Newton2016,Newton2018}. We used the methods presented in \citet{Irwin2011,Newton2016,Newton2018}. We searched periods ranging from 0.01--1000 days using a periodogram analysis. We found all periods are consistent with their previously published values. 

We used photometric observations from the Transiting Exoplanet Survey Satellite (\textit{TESS}) for one target, GJ~1111. This target was observed during Sector 21 from UT 2020 January 21 - UT 2020 February 18. We used the Pre-Search Data Conditioning Simple Aperture Photometry (PDCSAP) two-minute cadence light curve shown in Figure \ref{fig:tess_lc}.

\section{Spectroscopic Activity and Stellar Rotation}\label{sec:methods}
\begin{figure}[ht]
\includegraphics[scale=.5,angle=0]{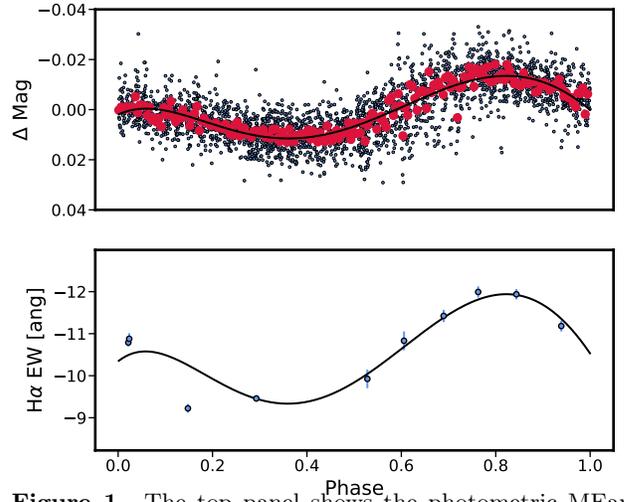}
\vspace{-1cm}
\hspace{-0.53cm}
\caption{The top panel shows the photometric MEarth data for G~7-34 phased to the stellar rotation period (blue points, black outlines) along with the spline fit to the MEarth data shown as the black solid line, and binned values plotted in red. The bottom panel show the $\ha$ equivalent width measurements phased to the rotation period of the star (blue points) and the best fitting model consisting of the spline fit to the photometry, with free parameters describing the amplitude, phase, and vertical offset (black solid line).  \label{fig:g734}}
\end{figure}

\begin{figure*}
\includegraphics[scale=.70,angle=0]{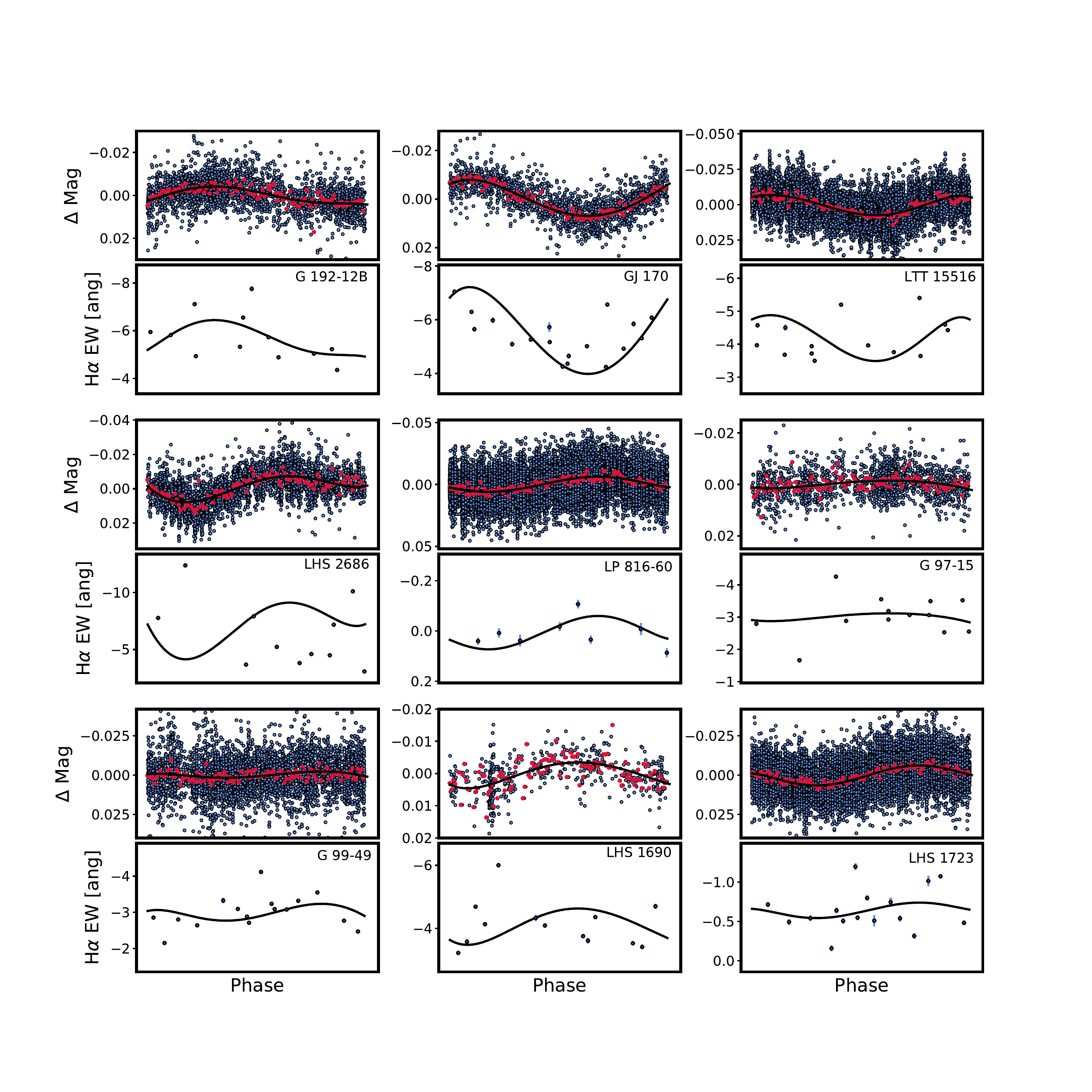}
\caption{The top panel in each pair shows the MEarth light curve for a given target phased to the stellar rotation period (blue points outlined in black) along with the spline fit to the MEarth data shown as the black solid line with binned values plotted in red. The bottom panels show the $\ha$ equivalent width measurements phased to the rotation period of the star (blue points) and the best fitting model consisting of the spline fit to the photometry, with free parameters describing the amplitude, phase, and vertical offset (black solid line). The name of the star is stated in the top right corner of the bottom panels. G 7-34 is not shown here, since it is plotted in Figure \ref{fig:g734}. \label{fig:all_stars}}
\end{figure*}

For each target we fit the null hypothesis, H$_{null}$, which is that the equivalent width is constant,

\begin{equation}
    {\rm H}_{null}(t)= C
\end{equation}

\noindent where C is a constant describing the zero-point offset of the EW data. We also fit an alternate hypothesis, H$_{alt}$, which is the data are correlated with the stellar rotational modulation,

\begin{equation}
    {\rm H}_{alt}(t) = C_{alt} + A~F(2\pi t/P  + \phi) 
\end{equation}

\noindent where C$_{alt}$ is a constant describing the zero-point offset,  $A$ is the amplitude of the EW variation, $F$ is a spline fit to the all the phased MEarth data for each target,  $P$ is the rotation period, and $\phi$ is the phase. We kept the period constant to the photometrically determined value, varying $A$, $C$ and $\phi$ to find the best-fitting parameters using maximum likelihood. Because H$_{null}$ and H$_{alt}$ are nested hypotheses, we used the F-test to determine whether H$_{alt}$ produced a significantly better fit to the data than H$_{null}$. The F-test is defined as:

\begin{equation}\label{eq:ftest}
    F = \frac{(RSS_{null} - RSS_{alt})/(dof_{null} - dof_{alt})}{RSS_{alt}/dof_{null}}
\end{equation}

\noindent where RSS$_{null}$ is the residual sum of squares of the fit to H$_{null}$, RSS$_{alt}$ is the residual sum of squares of the fit to H$_{alt}$, and dof$_{null}$ and dof$_{alt}$ are the degrees of freedom for H$_{null}$ and H$_{alt}$ respectively. The results from the F-test indicated that the null hypothesis (i.e. the variation in $\ha$ is not correlated with stellar rotation) is favored in all cases except G~7-34. For G~7-34, we observed that as the $\ha$ decreases (increased emission), the stellar brightness increases. We show the fit to H$_{alt}$ for G~7-34 in Figure \ref{fig:g734} (top panel). The F-test value for this star is 19.98 which is at the 98.3\% confidence level. We show the fit to H$_{alt}$ for the other stars in Figure \ref{fig:all_stars} and provide the F-test statistic and confidence levels for all targets in Table \ref{tab:stellar_params}. We test whether outlier $\ha$ measurements, likely due to enhancement from stellar flares, are affecting our results for the stars that do not show $\ha$ varying in phase with stellar rotation. To test this, we chose two stars LHS 1690 and G 170 and removed 3$\sigma$ outliers from the mean from each time series, fit H$_{null}$ and H$_{alt}$, and computed the F-test statistic. We found that even with outliers removed, the F-test was insignificant and thus we conclude the same result; the dominant variation on the other 9 stars is not originating from roughly constant emission that is modulated by spots rotating into and out of view.

\subsection{G 7-34 is a Member of AB Doradus}
G 7-34 was different from the other targets in that its $\ha$ emission correlated with rotational phase. We investigated the galactic kinematics of G 7-34 using the BANYAN $\Sigma$ Tool \citep{Gagne2018}, which takes as inputs the coordinates, radial velocity, proper motion in right ascension, $\mu_{\alpha}$, and declination, $\mu_{\delta}$ and parallax of the star to examine whether G~7-34 is associated with any young stellar associations. We used parallax = 68.55 $\pm$ 0.08 mas, $\mu_{\alpha}$ = 133.97 $\pm$ 0.15, $\mu_{\delta}$ = -377.25 $\pm$ 0.10 mas~yr$^{-1}$. We obtained these values from \textit{Gaia} Data Release 2 \citep{GaiaDR22018}. We measured the radial velocity by cross-correlating each high-resolution TRES spectrum of G 7-34 with a TRES spectrum of Barnard's Star and determined the average radial velocity value. For more details on this method, please see \citet{Winters2020}. We found the average radial velocity to be 15.06 $\pm$ 0.33 km~s$^{-1}$. We found, with 99.99\% confidence, that G~7-34 is associated with the AB Doradus Moving Group, which has an age of 149 Myr \citep{Zuckerman2004}. After we performed this analysis, we learned that its membership in the AB Doradus moving group was previously identified in \citep{Bell2015}. As this star is young, it is likely over-luminous leading to a modest overestimation of its mass determined using the absolute K$_s$ magnitude and relations presented in \citet{Benedict2016}. We also checked the colors of G~7-34 to ensure they were consistent with the color magnitude diagram for stars of this age: using the current literature, we were unable to compile a robust statistical sample of bonafide members of the AB Doradus Moving group with spectral types ranging from (M0-M9)V so we used the Pleiades Cluster which has a comparable age of 125 Myr \citep{Stauffer1998}. We used the V$_J$ and I$_{KC}$ band magnitudes of V$_J$ = 13.84 and I$_{KC}$ = 10.75 presented in \citet{Riedel2014}. After we adjusted to the distance modulus of the Pleiades (m - M) = 5.53 \citep{Stauffer1998} and accounted for reddening E(V$_J$-I$_{KC}$) = 0.06 \citet{Stauffer1987}, we found that V = 18.67 and (V$_J$-I$_{KC}$ )= 3.15. We used the color magnitude diagram presented in Figure 2 from \citet{Stauffer1998} and these values, found G~7-34 to be consistent with other Pleiades members. We found the rotation period of 0.37 days is consistent with the measured rotation periods of Pleiades members of a similar mass using the data presented in \citet{Redbull2016}; the rotation periods for stars with a similar V-K$_s$ = 5.65 to G 7-34 have rotation periods spanning (0.1-1.0) days. This star is younger than the other 9 stars in this study and it is the only star that displays $\ha$ variability in phase with the stellar rotation period. It is not immediately clear why the dominant source of $\ha$ variability on this young star is different than older active stars.

\section{High-Cadence Observations}\label{sec:HC}
\begin{figure*}
\includegraphics[scale=0.90,angle=0]{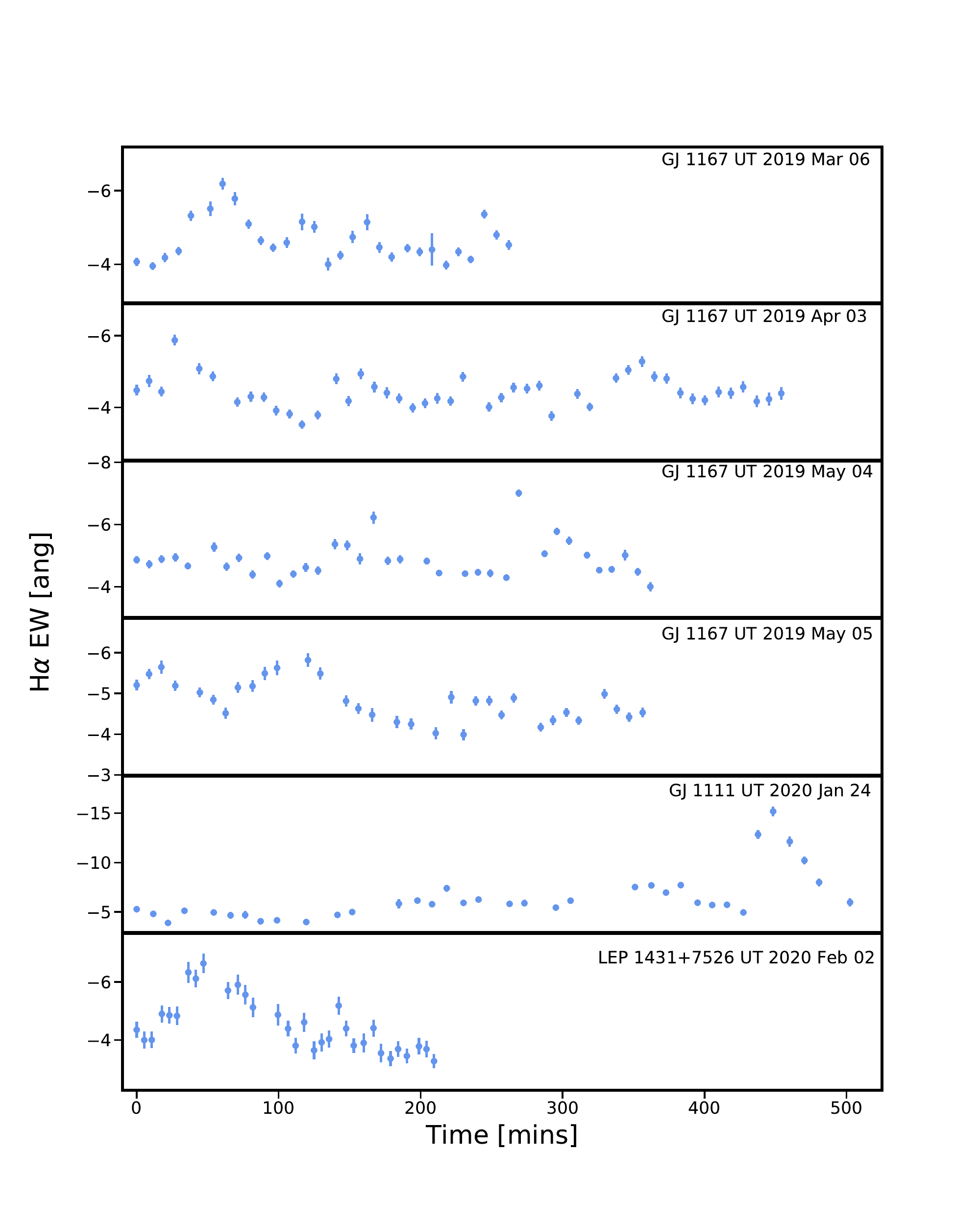}
\caption{Each panel displays the time-series of $\ha$ equivalent width measurements obtained during high-cadence observations over the course of one observing night. The top four panels show the four nights obtained for GJ~1167, followed by GJ~1111 and the bottom panel shows LEP~1431+7526. \label{fig:HC_obs}}
\end{figure*}

Our results indicated that for 9 out of 10 targets the dominant source of variability in $\ha$ is not regions of roughly constant emission that are modulated by spots rotating into and out of view. Nonetheless, the $\ha$ is clearly varying. With this null result, we probed further into determining the timescale of variability for $\ha$ as this may point to the possible mechanism responsible for the variation. We proceeded to obtain high-cadence observations of GJ~1167, GJ~1111, and LEP~1431+7526 as described in Section \ref{sec:spec_obs}. The resulting time series of $\ha$ measurements are shown in Figure \ref{fig:HC_obs}.

We explored the timescales of variability using two methods, an autocorrelation function (ACF) and a Gaussian process (GP). 

\subsection{Autocorrelation Function}
\begin{figure*}
\includegraphics[scale=0.80,angle=0]{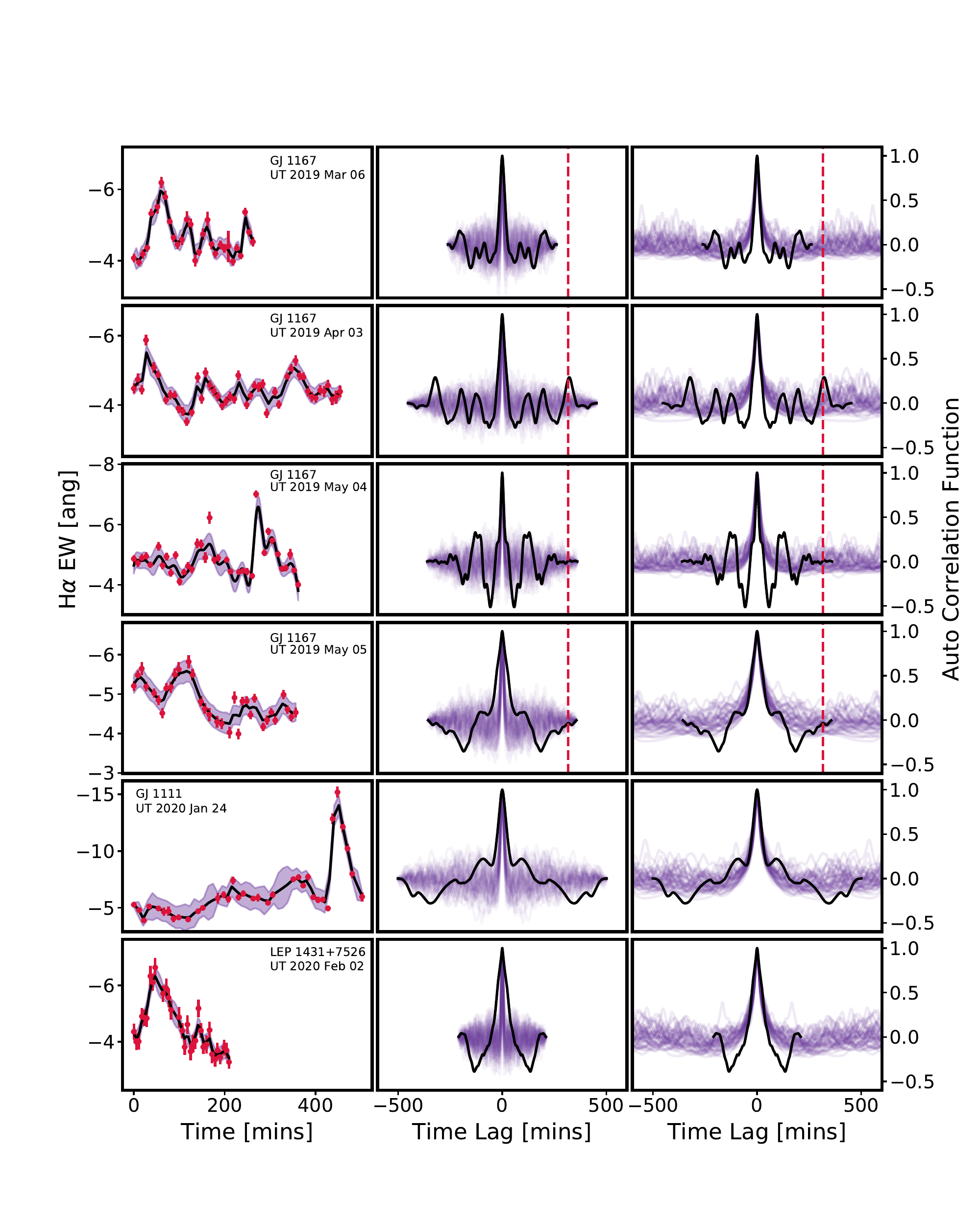}
\caption{Left column displays the time-series of $\ha$ equivalent width measurements obtained during high-cadence spectroscopic observations. The star name and the date and time of the observation are listed in the upper right. The Gaussian process we used to model the data is shown as the black curve and the one sigma uncertainty in the model is shown as the light purple shaded regions. The middle column shows the autocorrelation function of the $\ha$ equivalent width measurements (black curve). The purple lines show the ACFs of the fake data sets consisting of randomly drawn EW values. The purple lines are narrower than the measured ACFs, indicating that we have detected a correlation in the values. The right column shows the ACFs of the fake time series of EW measurements consisting of flares injected into the time series at random times with the same decay timescale as measured by the FWHM shown in Table \ref{tab:params} (purple curves). The black curve is the measured ACF of the real data, and is the same as shown in the middle column. We show the rotation period of GJ~1167 as the red dashed line; the rotation periods of the other two targets are longer than the durations of the respective data sets \label{fig:ACF}}
\end{figure*}

We computed the autocorrelation function for each high-cadence $\ha$ time series and plotted these in the middle and left columns of Figure \ref{fig:ACF}. We measured the full width at half maximum (FWHM) of the ACF as a probe for the correlation timescale of the data. We determined the point halfway between the global minimum and global maximum values of the ACF. We then mirrored the ACF so that is symmetric about zero. The FWHM is the difference between the time lags at which the halfway point of the ACF occurred. We determined the uncertainty on each FWHM measurement using the bootstrap method. We perturbed each EW measurement using 1000 Gaussian deviates with the standard deviation set to the error bar of the measurements. We then calculated the standard dedviation of the FWHM estimate from those samples taking that to be the uncertainty in the FWHM measurement. We present the FWHM measurements and their uncertainties in Table \ref{tab:params}. We found the FWHM values range from 30--70 minutes. 

To test whether we detected a significant correlation, we computed the ACF for 500 randomly distributed $\ha$ equivalent width measurements with timestamps equal to those of the true data sets. The EW values were drawn from a normal distribution with mean and standard deviation equal to those of each nightly observation. We found that the ACF that we measured for the actual data is substantially wider than the ACFs derived from the fake data sets (see Figure \ref{fig:ACF}). We concluded that the correlation we detect using an ACF is significant. 

\subsubsection{Autocorrelation Function of Simulated Flare Time-series}

Because the correlation timescales as deduced from our measurement of the FWHM of the ACFs and were similar to the typical duration of flares, and because flares are known to result in elevated $\ha$ emission, we considered whether flares provide a plausible explanation for the correlated variability we observed.  We proceeded to test this idea by creating mock data sets as follows.  We used the flare template from \citet{Davenport2014}, where each flare is characterized by a time of peak flux, the full-width at half maximum timescale of the flare, and the amplitude of the flare. The decay phase of the flare in the template is parameterized as the sum of two exponentially decaying components, one describing the fast, steep decay and the other describing the slower more gradual decay. The fast decay timescale dominates the flare duration. We injected 10 flares into 800 minutes of data with fast decay phase timescales ranging from 30--70 minutes following the distribution of FWHM values obtained with the ACF. For a given flare injected light curve, each flare had the same decay phase timescale within the 30--70 minute range and a randomly assigned amplitude and time of peak flux. We employed boxcar integration on the mock data sets to simulate and match the integration times of our spectroscopic observations for each star. We then computed the ACF and measured the FWHM of the ACF for each mock data set. We found that the measured ACF FWHM values are within 5\% of the input decay time scale. We concluded from this test, that flares can reproduce the correlated $\ha$ variability we observed in our high cadence time series. We show the ACFs of the mock data sets in the right panel of Figure \ref{fig:ACF}.

\begin{figure}[ht]
\includegraphics[scale=.6,angle=0]{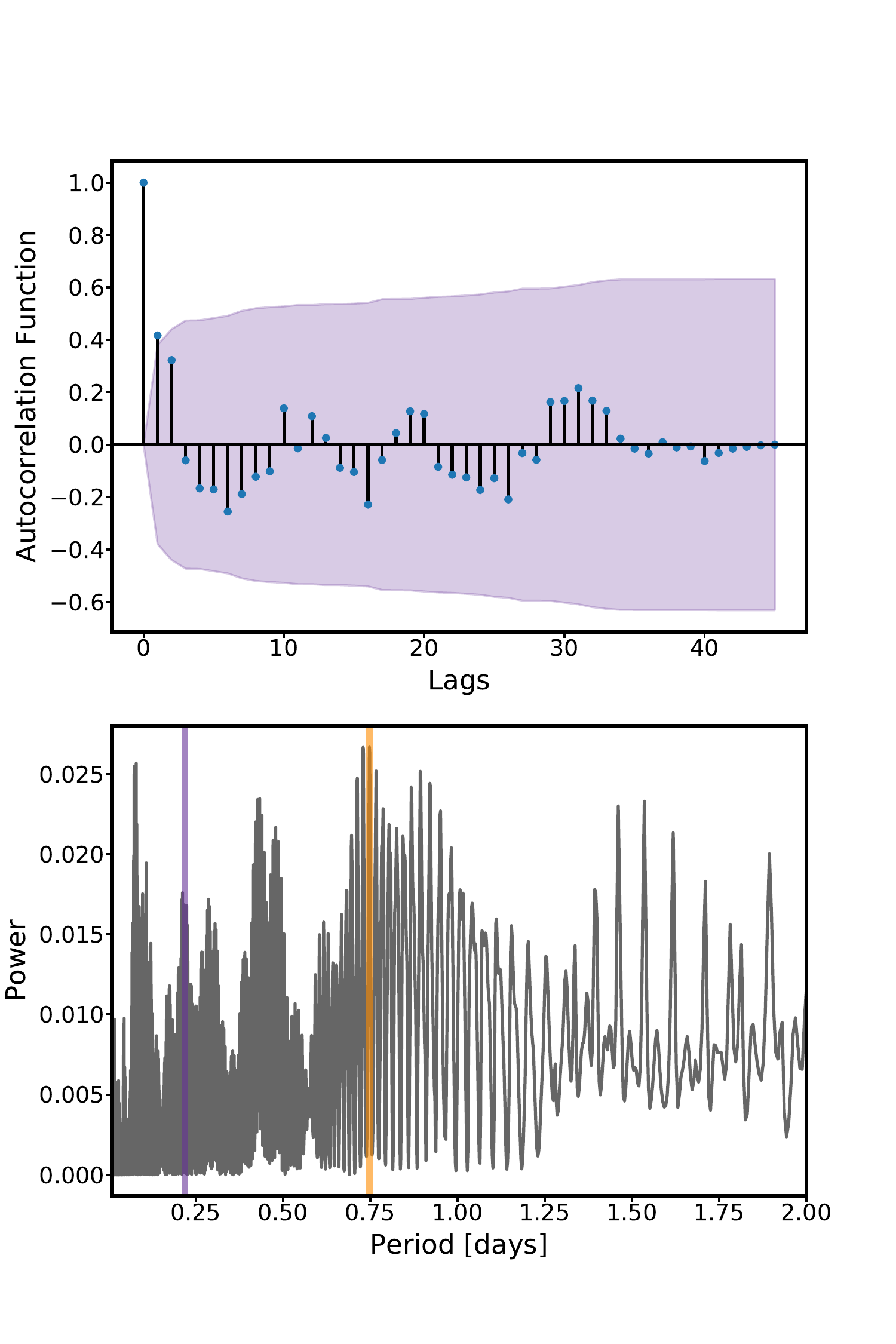}
\vspace{-1cm}
\hspace{-0.53cm}
\caption{The top panel shows the auto-correlation function for GJ 1167 on UT 2019 April 3. The purple contour shows the 99.7\% significant level interval. The bottom panel shows the Lomb-Scargle periodogram  of the combined four nights of GJ 1167. The purple line denotes the rotation period of GJ 1167 at 0.22 days. The orange line represents the peak of the periodogram which occurs at 0.74 days.  \label{fig:sig_gj1167}}
\end{figure}

We note that some of the ACFs of the real data show side lobes, notably GJ 1167, for which the ACF of the second night, UT 2019 April 3, displays a prominent side lobe at the rotation period of 316.8 minutes. We use the python package \textit{Statsmodels}  \citep{seabold2010statsmodels}  to compute the ACF and 99.7\% significant level interval for the second night of GJ 1167 observations. In Figure \ref{fig:sig_gj1167}, we plot the ACF as well as the 99.7\% significant level interval. From this, we conclude that the side lobes in the ACF are not significant. Furthermore, we see in Figure \ref{fig:ACF} that side lobes appear often in the simulated ACFs even though periodic phenomena are not present in the mock flare data sets as the flare times are chosen randomly. Additional evidence that the side lobes observed in the ACF of the real data are not significant is that they are similar or smaller in size in comparison to those observed in the mock data. These side lobes likely result from the time offset of the individual, relatively large flares. In addition, we conducted a periodogram analysis on the four full nights of data for GJ 1167 finding no significant peaks at the rotation period or any of its harmonics. We show the results of the periodogram in Figure \ref{fig:sig_gj1167}.

\subsection{Gaussian Process}
We probed the timescale upon which the data are correlated using a Gaussian process and the Matern 1/2 kernel, which is given by:

\begin{equation}\label{eq:gp}
    k(t) = \eta^2 {\rm exp}\left( \frac{-t}{\tau} \right) + \sigma^2 
\end{equation}

\noindent where $\tau$ is the time scale of the correlations, $\eta^2$ is the variance of the correlations, and $\sigma^2$ is a white noise term.  We used maximum likelihood estimation to determine the hyper-parameters $\tau$, $\eta$, and $\sigma$.  We used {\sc scikit learn} \citep{scikit-learn} to implement the GP and for the parameter estimation. In Figure \ref{fig:ACF} we show the GP fit for each high-cadence observation.  We chose the Matern 1/2 kernel because it describes covariances observed in fast rise and exponential decay problems. Given results from the injected flare ACF analysis in \S~\ref{sec:inject}, we posited that the correlation timescale we observed in $\ha$ is related to stellar flares which show an exponential decay in their decay phase and can enhance chromospheric activity indicators such as $\ha$ \citep{Kowalski2009,Kowalski2013,Fuhrmeister2018,DiMaio2020}. In this way, $\tau$ is probing the e-folding timescale of the flare decay.  We used bootstrap methods to determine the uncertainties in $\tau$. We found that the time scale of variability for the high-cadence data sets range from 20 -- 45 minutes. We show the best fit $\tau$ values and their bootstrap uncertainties in Table \ref{tab:params}.  We took these values of $\tau$ to be the true timescale of variability on the star instead of those values measured by the FWHM of the ACF. 

We also determined the time scale, $\tau$, of 100 mock flare data sets for each night of observations for GJ~1167, GJ~1111, and LEP~1431+7526 using this GP analysis. We find that if we tune the duration of the injected flare data sets so that the average output ACF FWHM value is consistent with the FWHM value we measured for the true data sets, the GP analysis returns times scales that are consistent with the values we measure with the GP values estimated for the true data sets. We provide the median and standard deviation $\tau$ of the 100 mock flare data sets for each night in Table \ref{tab:params}. 

\begin{center}
\begin{deluxetable*}{llccc} \label{tab:params}
\tablehead{ 
\colhead{Star} & 
\colhead{Date} &
\colhead{GP $\tau$} &   
\colhead{ACF FWHM} &
\colhead{GP $\tau$ Injected} \\
\nocolhead{yup}  & 
\colhead{UT} &
\colhead{minutes} &   
\colhead{minutes} &
\colhead{minutes}}
\startdata
GJ~1167 & 2019 Mar 06   & 19.4$\pm$ 2.8 & 34.5 $\pm$ 4.2 & 22.5  $\pm$ 9.5 \\
GJ~1167 & 2019 Apr 03   & 17.7$\pm$ 2.5 & 34.5 $\pm$ 4.6 & 22.5 $\pm$ 9.5 \\
GJ~1167 & 2019 May 04   & 20.0$\pm$ 2.7 & 40.3 $\pm$ 3.8 & 25.4  $\pm$ 7.6 \\
GJ~1167 & 2019 May 05   & 51.3$\pm$ 8.0 & 69.1 $\pm$ 4.8 & 46.6  $\pm$ 8.0\\
GJ~1111 & 2020 Jan 24   & 33.0$\pm$ 2.5 & 57.6 $\pm$ 5.4 & 36.2  $\pm$ 14.2\\
LEP~1431+7526 & 2020 Feb 02   & 56.3 $\pm$ 8.2 & 60.1 $\pm$ 8.7 & 45.4 $\pm$ 9.0 \\
\enddata
\caption{GP $\tau$ denotes the correlation timescale of the high cadence data sets as determined by the Gaussian process regression. ACF FWHM denotes the measured full-width at half maximum of the Auto-correlation function. GP $\tau$ Injected denotes the median correlation time  scale of the fake time series of EW measurements consisting of randomly injected flares with the same decay timescale as is measured by the ACF FWHM}
\end{deluxetable*}
\end{center}


\subsection{ARIMA Modeling of GJ 1167}

As a separate path to explore the correlations in the $\ha$ time series, we used the AutoRegressive Integrated Moving Average (ARIMA) on the full four night data set of GJ 1167. ARIMA is useful for modeling stationary time series and is often employed as a forecasting tool \citep{Feigelson2018}. ARIMA can be applied to data only if they are evenly spaced, so we first binned the data for GJ 1167 into bins of 8.6 minutes. We then used the Python Package \textit{Statsmodels} \citep{seabold2010statsmodels} to implement ARIMA. We find that an ARIMA model of (1,0,0) is favored to describe the GJ 1167 time series, with an Akaike information criterion equal to 202. ARIMA(1,0,0) describes a damped random walk, which is sometimes termed an Ornstein-Uhlenbeck process \citep{Uhlenbeck1930}. This describes a time series that has a tendency to return to its mean value, and the strength of the returning attraction is proportional to the size of the deviation from the mean. This is consistent with our physical picture, namely that $\ha$ EW returns to its quiescent value after a flare. The ARIMA framework, however, does not quantify the return time scale, but we derive this time scale using the auto-correlation and Gaussian processes methods described earlier.

\subsection{Simultaneous TRES and \textit{TESS} Observations}\label{sec:inject}
For one high-cadence target, GJ~1111,  we gathered the spectroscopic time series described above at the same time the star was also observed by \textit{TESS}. The TRES observations cover 8.37 hours of the 11.04 hour rotation period. 

\begin{figure*}
\includegraphics[scale=0.70,angle=0]{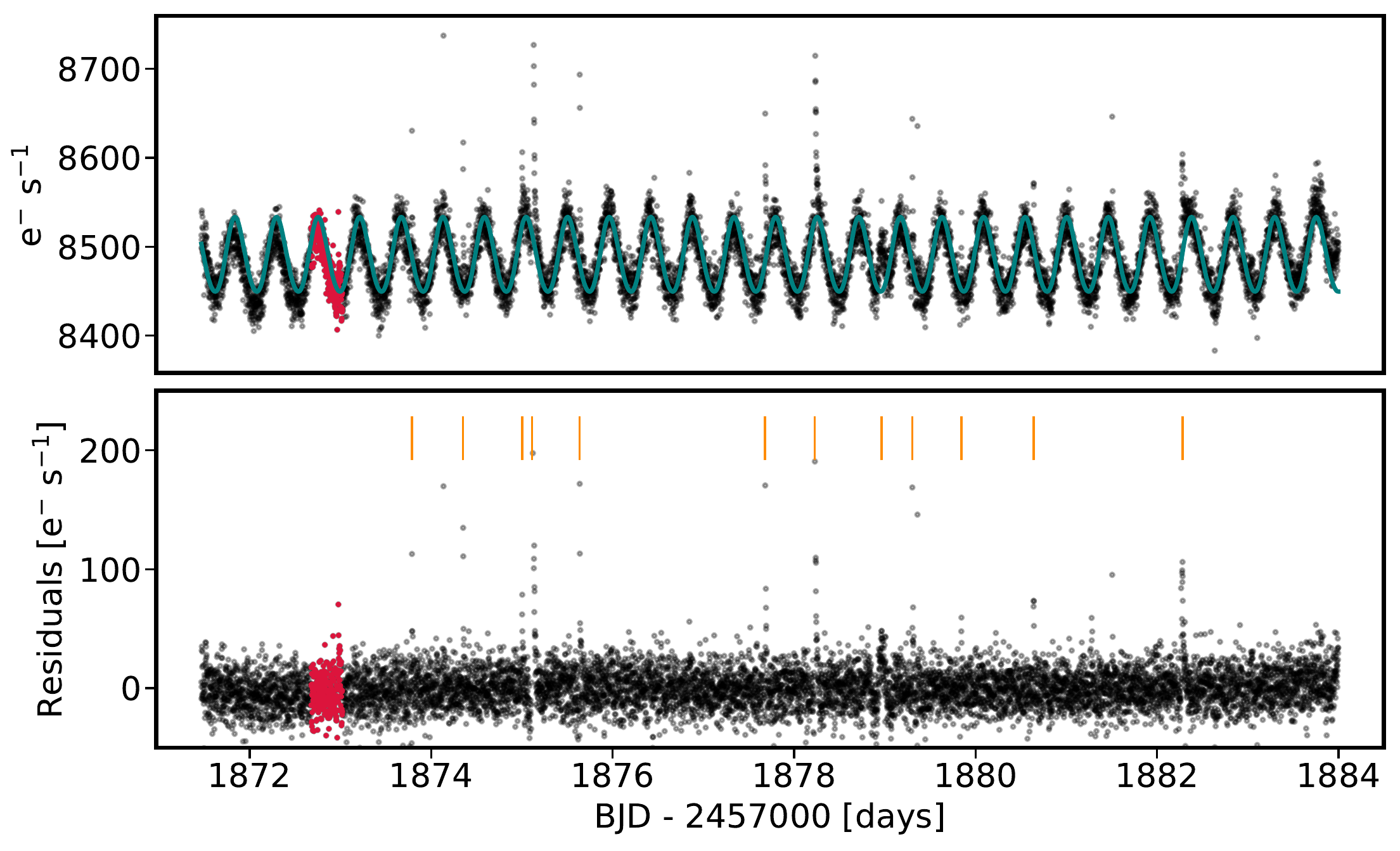}
\caption{The top panel shows the \textit{TESS} PDCSAP light curve of GJ~1111 during the \textit{TESS} orbit where simultaneous spectroscopic observations were obtained. The red points indicate the times during which the simultaneous TRES observations were gathered. The teal curve is the GP model we used to model the light curve. The bottom panel shows the residuals with flares found by our flare finding algorithm shown with an orange tic mark.\label{fig:tess_lc}}
\end{figure*}

We used the \textit{TESS} data to probe whether the variation of $\ha$ observed in Figure \ref{fig:HC_obs} can be attributed to stellar flares. To detect stellar flares, we first removed the photometric modulation due to stellar spots rotating into and out of view. We proceeded as described in \citet{Medina2020}: we first modeled the light curve using a Gaussian process taking the form of a combination of two simple harmonic oscillators from the python package {\sc exoplanet} \citep{exoplanet:foremanmackey17}.  We then searched for flares, which are defined as 3 consecutive 3$\sigma$ positive deviants. Our algorithm is able to detect flares (30\% completeness; see \citet{Medina2020}) with energies greater than  4$\times$10$^{30}$ ergs in the \textit{TESS} bandpass. We detected 24 flares total in the full 27 day, two-orbit \textit{TESS} light curve. The detected flares range from 4.3$\times$10$^{30}$ to 8.2$\times$10$^{31}$ ergs. Our flare finding algorithm did not detect any flares during the overlapping timestamps. We show the detected flares for \textit{TESS} orbit which overlapped with the spectroscopic observations in Figure \ref{fig:tess_lc}. 

\begin{figure*}
\includegraphics[scale=0.70,angle=0]{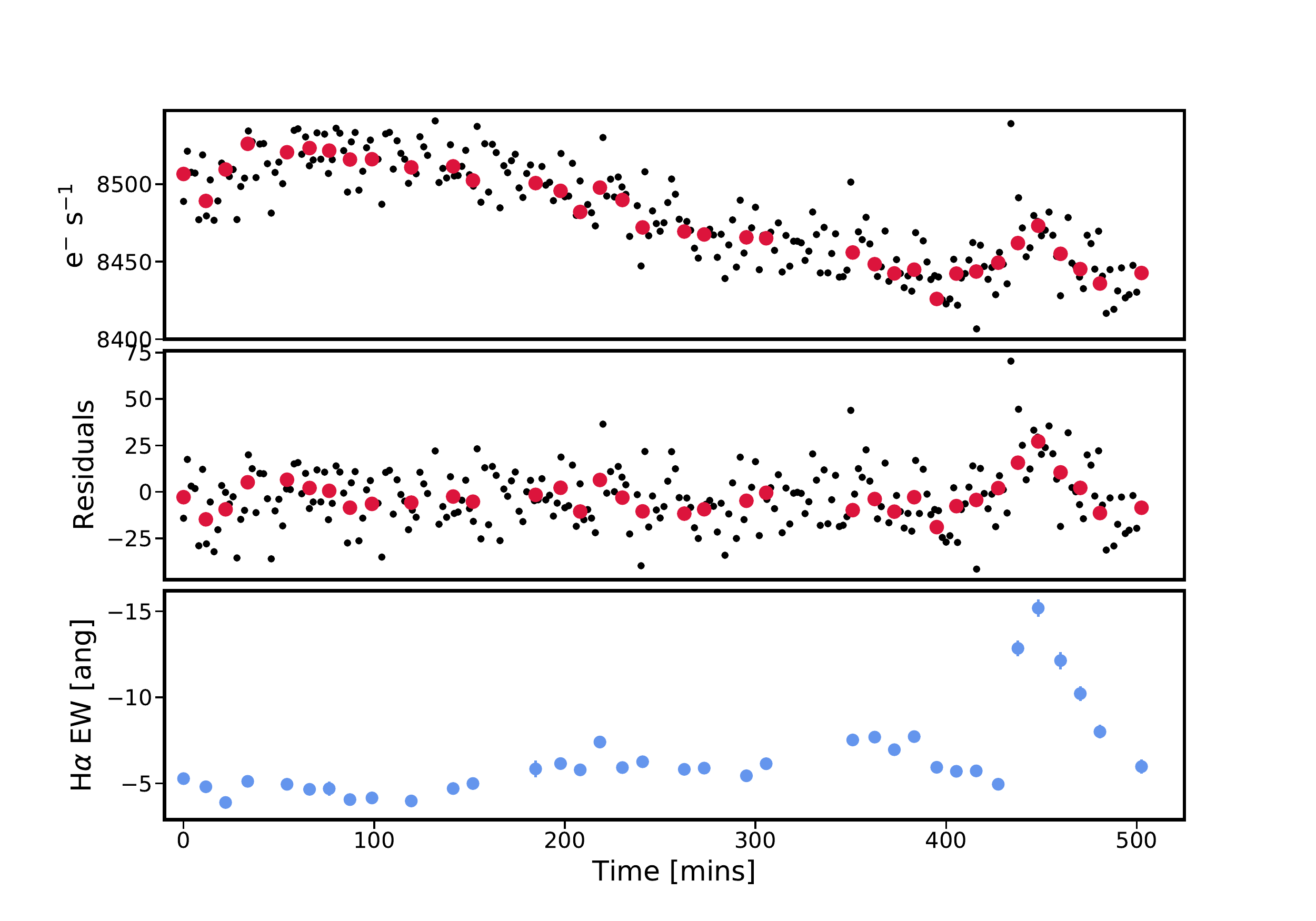}
\caption{The top panel shows \textit{TESS} PDCSAP light curve (black points) with red points showing the \textit{TESS} light curve binned to match the times and durations of exposures of each TRES spectrum. The middle panel shows the residuals of the \textit{TESS} light curve of GJ~1111 after the subtraction of the Gaussian process model we used to detrend the data. The bottom panel shows the time-series of the $\ha$ EW measurements. \label{fig:sim}}
\end{figure*}

In Figure \ref{fig:sim}, we show the \textit{TESS} PDCSAP light curve, the detrended light curve and $\ha$ equivalent width measurements during the time of simultaneous observations. Although our algorithm did not not detect a flare during the simultaneous observations, we did observe a positive excursion in both the detrended and PDCSAP \textit{TESS} light curves that aligns with the peak of the $\ha$ equivalent width time series (Figure \ref{fig:sim}). We believe this feature observed in both the TRES and \textit{TESS} data sets may be a low-energy flare that is below the sensitivity of our algorithm. To test whether the $\ha$ time series and the \textit{TESS} photometric time series are correlated, we first binned the \textit{TESS} data to match the exposure times of each TRES spectrum. We then calculated the cross-correlation of the $\ha$ EWs and the binned \textit{TESS} data. We tested the correlation of the $\ha$ time series and the binned PDCSAP light curve, as well as, the detrended light curve by calculating the Spearman's Rank and Pearson Correlation coefficients respectively as well as their associated P-values. The P-value indicates the probability of an uncorrelated data set producing the same coefficient value. We found the Pearson Correlation coefficient = 0.700, p-value = 2.04$\times$10$^{-6}$ for the detrended light curve and the $\ha$ time series and the Spearman's Rank Correlation coefficient = 0.676, p-value = 5.93 $\times$10$^{-6}$ for the correlation between the PDCSAP light curve and the $\ha$ time series. 

We examined the correlation between the detrended and undetrended PDCSAP light curve and the $\ha$ time series with the potential flare excluded (0 $<$ t $<$ 400 minutes) from the analysis to understand the effect of the flare on the correlation. We found the Pearson Rank Correlation coefficient = -0.01, p-value = 0.91 and  Spearman's Rank Correlation coefficient = 0.78, p-value = 1.59$\times$10$^{-6}$ respectively for the detrended and undetrended light curves.  For completeness, we also computed the two correlation coefficients for the region only including the flare for times greater than 400 minutes. We find the values of -0.82, p-value = 7.34$\times$10$^{-4}$, and -0.63, p-value = 6.71$\times$10$^{-3}$ for the Pearson and Spearman’s Rank Correlations respectively. Finding correlations between the $\ha$ times series with the exclusion of the potential flare as well as with the potential flare only portion of the light curve suggests that $\ha$ variability may include contributions from fixed magnetic phenomena rotating into and out of view as well as flares.

\section{Discussion and Conclusion}\label{sec:conclusion}
We used measurements of $\ha$ equivalent widths as a function of time to examine the timescales upon which this chromospheric activity indicator varies on fully convective M-dwarfs with masses between 0.1 and 0.3 solar masses. We studied 13 active stars with rotation periods spanning from 0.21 to 92 days. 
We first examined whether the dominant source of variation in $\ha$ is due to roughly constant emission from localized magnetic phenomena such as stellar spots or plages and thus varies in phase with stellar rotation period and, if so, whether the correlation shows a mass dependence. Only one star, G 7-34, showed variations that are correlated with the stellar rotation period; as the stellar brightness increases, the star shows increased $\ha$ emission. We found, through a kinematic analysis, that this star is a member of the AB Doradus moving group and thus younger than the other nine stars. G 7-34 is the most active star in our sample showing the greatest amount of $\ha$ in emission and also has the largest photometric amplitude. We posit that the asymmetric distribution of potentially larger stellar spots could be continuously emitting which could account for the variation of $\ha$ being in phase with the stellar rotation. The remaining nine stars showed $\ha$ variability, but the variation is inconsistent with the rotational phase of the star. These stars may have more numerous, small spots that display more sporadic chromospheric emission at all rotational phases with flares erupting randomly from these spots and fading away. This stands in contrast to findings for more massive inactive M-dwarfs, which show that chromospheric activity indicators like Ca II H+K and $\ha$ trace the stellar rotation periods  \citep[e.g.][]{Suarez2016,Fuhrmeister2019,Shofer2019}. These previous findings indicated that the dominate source of activity variability originates from relatively constant emission from localized manifestations of the magnetic field through active regions in the chromosphere.


We obtained six nights of high-cadence observations for three stars. We found that each star shows correlated variability on timescales of 20-45 minutes, which is much shorter than the rotation period.   \citet{Hawley2014} showed that the flare decay timescale distribution for the active mid M dwarf GJ~1243 peaks ranges from 10--100 minutes. The variability timescale that we observed is consistent with this timescales. 


From simultaneous spectroscopic and photometric monitoring (with \textit{TESS}) of GJ 1111, we found that as the stellar brightness increased, the amount of $\ha$ in emission also increased. This correlation may be due to low-energy flaring events, which is in agreement with results from \citet{Lee2010} and \citet{Bell2012} in regard to their findings for stars with masses below the convective boundary. In those studies, they noted that $\ha$ is variable, but do not determine whether the variability is correlated in time. The enhanced $\ha$ emission caused by low-energy flares may be masking the underlying localized $\ha$ activity observed in inactive stars with $\ha$ in absorption \citet{Suarez2016,Fuhrmeister2019} where the flare rate for these stars can be reduced 6 orders of magnitude or more \citep{Medina2020}. 

The question of how fully convective M-dwarfs generate and sustain their magnetic fields is of intrinsic interest and bears upon the habitability of their orbiting planets. These stars provide the only near-term opportunity to characterize the atmospheres of terrestrial exoplanets. As such, it is essential that we understand the past and present stellar radiation environment, including the timescales upon which magnetic phenomena are occurring. Knowing these timescales is useful for both our understanding of the manifestations of the magnetic dynamo and also for radial velocity surveys. The major inhibitor to detecting planetary signals in radial velocity surveys is disentangling them from the underlying stellar variability \citep{Robertson2014a, Robertson2014b, Robertson2015}. 

In this paper, we explored the timescale of variability for the chromospheric activity indicator $\ha$. Our results suggest that stellar flares present a feasible mechanism to describe $\ha$ variability on active mid-to-late M dwarfs as opposed to roughly constant emission from active regions, such as stellar spots or plages rotating into and out of view like as has been observed on inactive early M-dwarfs. Whether this relationship extends to inactive mid-to-late M dwarfs has not yet been studied extensively. In a future study, we intend to examine the link between low-energy flaring behavior and $\ha$ variability further by obtaining simultaneous spectroscopic and \textit{TESS} observations for a larger sample of stars.

\acknowledgments{We would like to thank the referee for a thoughtful review that improved the manuscript. We thank David Latham, Gilbert Esquerdo, Perry Berlind, and Michael Calkins for scheduling and collection of the TRES data. This work is made possible by a grant from the John Templeton Foundation. The opinions expressed in this publication are those of the authors and do not necessarily reflect the views of the John Templeton Foundation. The MEarth Team gratefully acknowledges funding from the David and Lucile Packard Fellowship for Science and Engineering (awarded to D.C.) and support from the National Science Foundation under grants AST-0807690, AST-1109468, AST-1004488 (Alan T. Waterman Award), and AST-1616624. This material is based upon work supported by the National Aeronautics and Space Administration under grant 80NSSC19K0635 in support of the \textit{TESS} Cycle 2 Guest Investigator program, and grant 80NSSC19K1726  issued through the XRP program. This paper includes data collected by the \textit{TESS} mission, which are publicly available from the Mikulski Archive for Space Telescopes (MAST). This work has made use of data from the European Space Agency mission Gaia (https://www.cosmos.esa.int/gaia), processed by the Gaia Data Processing and Analysis Consortium (DPAC, https://www.cosmos.esa.int/web/gaia/dpac/consortium). Funding for the DPAC has been provided by national institutions, in particular the institutions participat-ing in the Gaia Multilateral Agreement. This work has used data products from the Two Micron All Sky Survey, which is a joint project of the University of Massachusetts and the Infrared Processing and Analysis Center at the California Institute of Technology, funded by NASA and NSF.}

\facilities{\textit{TESS}, MEarth, FLWO:1.5m (TRES)} 

\software{{\sc celerite} \citep{Foreman-Mackey(2017)}, {\sc exoplanet} \citep{exoplanet:exoplanet}, {\sc PYMC3} \citep{exoplanet:pymc3}, {\sc python}}

This research made use of {\sc exoplanet} \citep{exoplanet:exoplanet} and its dependencies \citep{exoplanet:astropy13, exoplanet:astropy18, exoplanet:exoplanet, exoplanet:foremanmackey17, exoplanet:foremanmackey18, exoplanet:kipping13, exoplanet:luger18, exoplanet:pymc3, exoplanet:theano}.
\clearpage
\bibliographystyle{aasjournal}
\bibliography{references}{}



\end{document}